\documentstyle[secnumtab]{fbssuppl}
\input{psfig.sty}

\title{Nuclear Reactions of Astrophysical Interest Involving Light
Nuclei}
\author{H. Oberhummer\thanks{{\it E-mail address}: ohu@ds1.kph.tuwien.ac.at},
W. Balogh, V.D. Efros\thanks{{\it Permanent address}:
Kurchatov Institute, Institute for General and Nuclear
Physics, SU--123182 Moscow, Russia}, H. Herndl,
R. Hofinger}
\institute{Institute of Nuclear Physics, Technical University Vienna, Wiedner
Hauptstr.~8--10, A--1040 Wien, Austria}

\begin{document}

\maketitle
\begin{abstract}
An introduction to nucleosynthesis, the creation of the elements in the
big bang, in interstellar matter and in stars is given. The two--step
process $^4$He(2n,$\gamma$)$^6$He and the reverse photodisintegration
$^6$He($\gamma$,2n)$^4$He involving the halo nucleus $^6$He
could be of importance in the $\alpha$--process in type--II supernovae.
The reaction rates for the above processes are calculated using three--body
methods and show an enhancement of more than three orders of magnitude
compared to the previous adopted value. Direct--capture calculations
give similar values for the above reaction rates.
Therefore, this method was also used to calculate the
reaction rates of the two--step processes $^6$He(2n,$\gamma$)$^8$He and
$^9$Li(2n,$\gamma$)$^{11}$Li and the reverse
photodisintegration of $^8$He and $^{11}$Li
that could be also of importance in the $\alpha$-process.
\end{abstract}

\section{Introduction}

One of the main driving forces for the evolution of our universe are
nuclear processes. These nuclear processes are responsible for
the energy production in stars as well as for the production of the elements
in our universe. The process of the creation of the elements is called
nucleosynthesis.
We have three different scenarios, where nucleosynthesis occurs: in the
big bang (primordial nucleosynthesis), in interstellar space (interstellar
nucleosynthesis) and in stars (stellar nucleosynthesis). The result
of nucleosynthesis is reflected today by the observed elemental abundances
in different objects (sun and other stars, planets, meteorites,
interstellar matter, \ldots).

The light elements from hydrogen to lithium with mass numbers $A=1$ to $A=7$
were mainly synthesized in primordial nucleosynthesis.
In interstellar nucleosynthesis primarily elements from lithium to
boron with mass numbers $A=6$ to $A=11$ were produced. Finally, all the heavier
elements
from carbon to uranium (in the astrophysical jargon these
elements are called metals) with mass numbers $A=12$ to $A=238$ were
created in stellar nucleosynthesis. Even though the heavier elements
amount to about only 1\% of the total observed abundance they are
obviously indispensable for our environmemt. Without stellar nucleosynthesis
our universe would be a boring place with the two gases hydrogen and
helium (with a negligible amount of lithium, beryllium and boron) and without
all the elements
necessary for the existence of life.

One of the essential inputs for the investigation of
different astrophysical scenarios for nucleosynthesis
are cross sections, astrophysical S--factors and reaction
rates of nuclear reactions as well as half lives and decay rates of
nuclear decays. These quantities are determined
using different experimental facilities (accelerators, reactors, \ldots).
However, in many cases it is not possible to obtain
the needed cross sections directly from experiments.
This is due to the fact that astrophysically relevant reactions take
place mainly at thermonuclear energies.
These energies are mostly well below
the Coulomb and/or centrifugal barrier, and in this region the
cross sections are often too small
for experimental determination. Furthermore,
in nucleosynthesis often radioactive nuclei are involved,
for which experimental information is only scarcely available.
For nuclei far--off stability often
experimental information is not attainable at all.

In the next section a short introduction to nucleosynthesis
is given emphasizing those astrophysical scenarios, where light
nuclei are involved and methods developed in the few--body field could
be applied. In Sect.~3 we show as an example for an application
of few--body calculations for astrophysically relevant nuclear
reactions the two--step process $^4$He(2n,$\gamma$)$^6$He.
This reaction involves also the halo nucleus $^6$He.
In the last section we calculate the reaction rate for this reaction as
well as other two--step reactions involving
halo nuclei, like $^6$He(2n,$\gamma$)$^8$He and
$^9$Li(2n,$\gamma$)$^{11}$Li, using the simpler direct--capture (DC) model.

\section{Nucleosynthesis}

In this section we only can give a short introduction to nucleosynthesis.
There are many textbooks where this subject can be found
in more detail (e.g.~\cite{rol88}).

Primordial nucleosynthesis took place as the cooling early universe reached
a temperature of about $10^9$\,K about three minutes after the big bang.
Before that time the temperature was high enough that formed deuterons
were immediately again destroyed through photodisintegration.
{}From primordial nucleosynthesis calaculations we can
deduce abundance ratios of the light stable
nuclides $^1$H, $^2$H, $^3$He, $^4$He, $^6$Li and $^7$Li. Even some
primordial abundance ratios may be altered somewhat by interstellar
and stellar nucleosynthesis, the agreement
between the abundances obtained from
primordial nucleosynthesis with the observed abundances is remarkable.

In interstellar nucleosynthesis the nuclides $^6$Li, $^7$Li,
$^9$Be, $^{10}$B and $^{11}$B are produced. In this scenario high--energy
(larger than about 1\,GeV)
cosmic rays hit interstellar matter and create through spallation processes
the above nuclei.

\begin{table}[htb]
\caption{\label{t1} Nuclear burning phases and corresponding temperatures}
\begin{center}
\begin{tabular}{lc}
\hline
Nuclear burning phase & Temperature ($T_{9}$)\\
\hline
Hydrogen burning & 0.01--0.04\\
Helium burning & 0.1--0.2\\
Carbon burning & 0.6--0.8\\
Neon burning & 1.2--1.4\\
Oxygen burning & 1.5--2.2\\
Silicon burning & 3--4\\
\hline
\end{tabular}
\end{center}
\end{table}

Stellar nucleosynthesis begins when through gravitational
contraction in a star the density and temperature in the
core gets high enough so that nuclear processes are ignited.
Stellar nucleosynthesis takes place in succesive burning phases.
When the fuel
of a preceding burning phase has been consumed, the
ash of this burning phase is the fuel for the next burning phase.
In the outer layers the previous burning phases still continue.
The different burning phases are hydrogen burning (pp--chain, CNO--cycle,
Ne--Na cycle, Mg--Al cycle),
helium, carbon, neon, oxygen and silicon
burning. In hydrogen burning helium is produced, in helium burning
carbon and oxygen are synthesized, in the remainig advanced
burning phases elements up to elements in the region of iron are produced.
In successive burning phases the temperature must get larger,
because of the increasing
Coulomb barriers of the fusing heavier nuclei (Table~\ref{t1}).

The stars in which hydrogen burning takes place are
named main--sequence stars, whereas the other
burning phases occur in red giants. Not
in all stars all the burning phases are ignited.
For stars with less than approximately 8 solar masses,
after helium burning, the outer part of
the star that is blown off by strong stellar winds is called a planetary
nebula, whereas
the core remains as a white dwarf. For
stars with more than about 8 solar masses all the
above cited burning phases take place. Finally,
after silicon burning the
core of the star consists of elements with mass numbers
in the region of iron. Then it is not possible
to gain energy anymore through nuclear fusion and the star
collapses. Through the rebounce effect after the implosion the stellar mantle
is ejected in a gigantic explosion, whereas the core remains as
a neutron star or black hole. This dramatic event is called
a supernova of type II\@ (a supernova of type I takes place
in a binary system consisting of two stars, where triggered through
mass accretion from the accompaning star the star
is disrupted completely).

There are astrophysical scenarios in which free neutrons play
a dominating role. Most elements heavier than iron have been created through
neutron capture in two processes: in the s--process occuring
in helium burning of red giants and in the r--proccess occuring in a
supernova of type II\@. The neutron flux in the s--process of helium
burning is comparetively low ($N_{\rm n} \simeq 10^{8}$\,cm$^{-3}$),
whereas in the r--process it is much higher
($N_{\rm n} \simeq 10^{20}$\,cm$^{-3}$).

\begin{figure}[htp]
\centerline{\psfig{file=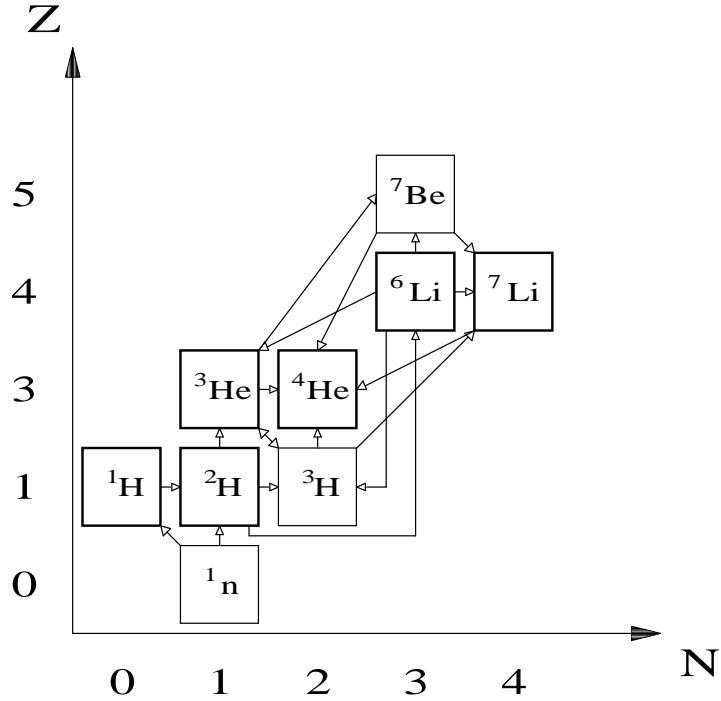,angle=270,width=18cm}}
\caption[]{\label{f1} Primordial nucleosynthesis in the Standard Big Bang
(SBB)}
\end{figure}

\begin{figure}[htp]
\centerline{\psfig{file=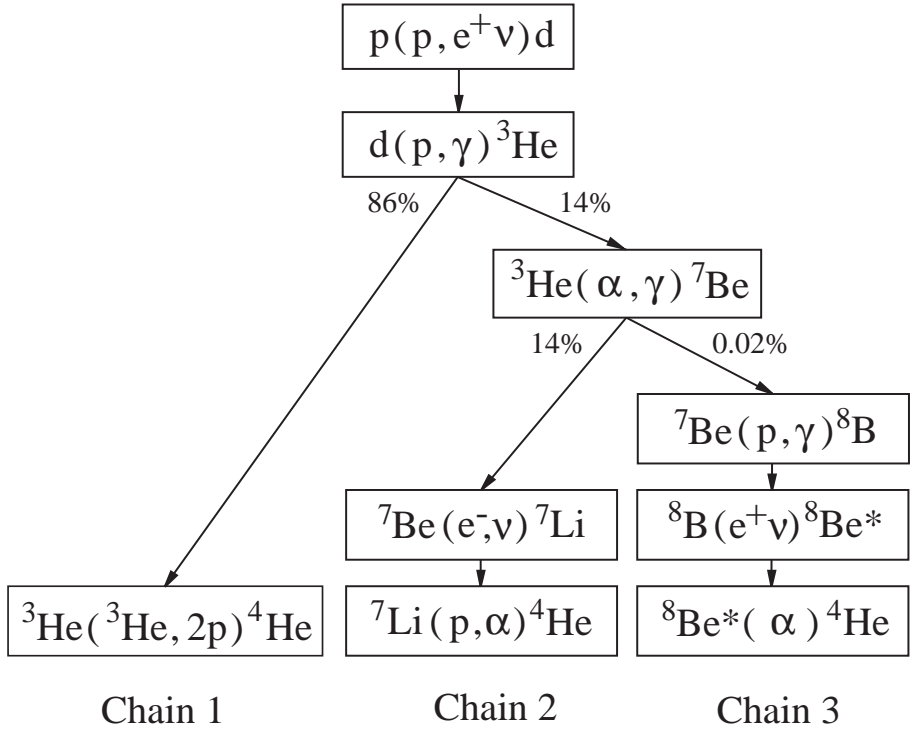,width=12cm}}
\caption[]{\label{f2} The pp--chain in hydrogen burning of main--sequence
stars}
\end{figure}

\begin{figure}[htp]
\centerline{\psfig{file=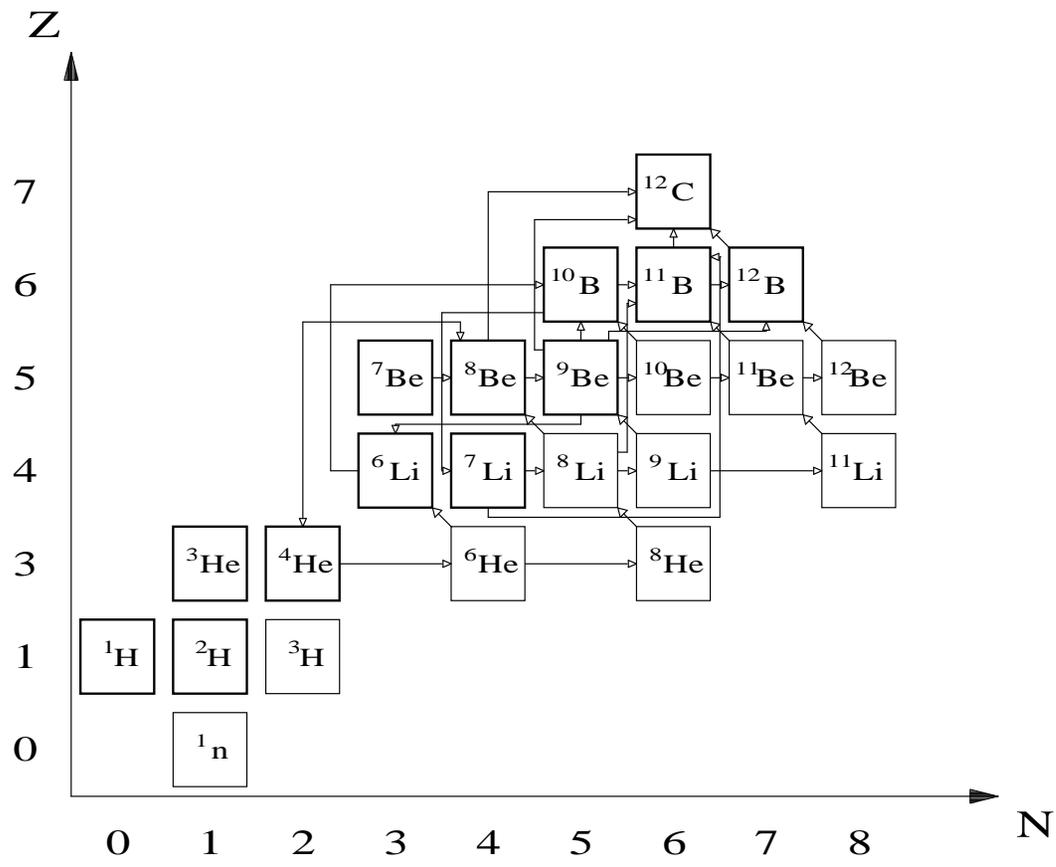,width=12cm}}
\caption[]{\label{f3} Additional reaction network in primordial nucleosynthesis
in the $\alpha$--process in
type II supernovae or the Inhomogenous Big Bang (IBB)}
\end{figure}

\begin{table}[htb]
\caption[Nuclear reactions for light nuclei]
{\label{t2} Nuclear reactions of astrophysical
interest for light nuclei up to $A=12$.
The left column lists the major reactions for nucleosynthesis
in the Standard Big Bang (SBB), additional
reactions in the pp--chain and helium burning.
The right column lists reactions which can be additionaly activated
in the $\alpha$--process or the Inhomogenous Big Bang (IBB).}
\begin{center}
\begin{tabular}{|lll||lll|} \hline
Target & Residual & Type of & Target & Residual & Type of  \\
Nucleus & Nucleus & Reaction & Nucleus & Nucleus & Reaction \\
\hline
\multicolumn{3}{|c||}{SBB} &
\multicolumn{3}{c|}{$\alpha$--process, IBB}\\
$^1{\rm n}$ & $^1$H & $\beta^-$ &
$^4$He & $^6$He & $(2{\rm n},\gamma)$ \\
$^1$H & $^2$H& $({\rm n},\gamma)$ &
$^6$He & $^8$He & $(2{\rm n},\gamma)$ \\
$^2$H & $^3$H& $({\rm n},\gamma)$, $({\rm d},{\rm p})$ &
$^6$He & $^6$Li & $\beta^-$ \\
$^2$H & $^3$He & $({\rm p},\gamma)$, $({\rm d},{\rm n})$ &
$^6$Li & $^{10}$B & $(\alpha,\gamma)$ \\
$^2$H & $^6$Li & $(\alpha,\gamma)$ &
$^7$Li & $^8$Li & $({\rm n},\gamma)$ \\
$^3$H & $^3$He& $\beta^-$ &
$^7$Li & $^{11}$B & $(\alpha,\gamma)$ \\
$^3$H & $^4$He & $({\rm d},{\rm n})$, $({\rm p},\gamma)$ &
$^7$Be & $^8$Be & $({\rm n},\gamma)$ \\
$^3$H & $^7$Li & $(\alpha,\gamma)$ &
$^8$He & $^8$Li & $\beta^-$ \\
$^3$He & $^3$H & $({\rm n},{\rm p})$ &
$^8$Li & $^9$Li & $({\rm n},\gamma)$ \\
$^3$He& $^4$He & $({\rm d},{\rm p})$, $({\rm n},\gamma)$ &
$^8$Li & $^8$Be & $\beta^-$ \\
$^3$He& $^7$Be & $(\alpha,\gamma)$ &
$^8$Li & $^{11}$B & $(\alpha,{\rm p})$ \\
$^6$Li & $^3$H & $({\rm n},\alpha)$ &
$^8$Be & $^9$Be & $({\rm n},\gamma)$ \\
$^6$Li & $^3$He & $({\rm p},\alpha)$ &
$^9$Li & $^9$Be & $\beta^-$ \\
$^6$Li & $^7$Li & $({\rm n},\gamma)$ &
$^9$Li & $^{11}$Li & $(2{\rm n},\gamma)$ \\
$^6$Li & $^7$Be & $({\rm p},\gamma)$ &
$^9$Be & $^6$Li & $({\rm p},\alpha)$ \\
$^7$Li & $^4$He & $({\rm p},\alpha)$ &
$^9$Be & $^{10}$Be & $({\rm n},\gamma)$ \\
$^7$Be & $^7$Li & $({\rm e}^-,\nu)$ &
$^9$Be & $^{10}$B & $({\rm p},\gamma)$, $({\rm d},{\rm n})$ \\
$^7$Be & $^4$He & $({\rm n},\alpha)$ &
$^9$Be & $^{12}$B & $(\alpha,{\rm p})$ \\
$^7$Be & $^7$Li & $({\rm n},{\rm p})$ &
$^9$Be & $^{12}$C & $(\alpha,{\rm n})$ \\
\cline{1-3}
\multicolumn{3}{|c||}{pp--chain} &
$^{10}$Be & $^{11}$Be & $({\rm n},\gamma)$ \\
$^1$H & $^2$H & $({\rm p},{\rm e}^{+}\nu)$ &
$^{10}$Be & $^{10}$B & $\beta^-$ \\
$^2$H & $^3$He & $({\rm p},\gamma)$  &
$^{10}$B & $^7$Li & $({\rm n},\alpha)$ \\
$^3$He & $^4$He & $(^3$He$,2{\rm p})$ &
$^{10}$B & $^{11}$B & $({\rm n},\gamma)$ \\
$^3$He & $^7$Be & $(\alpha,\gamma)$ &
$^{11}$Li & $^{11}$Be & $\beta^-$ \\
$^7$Be & $^8$B & $(p,\gamma)$ &
$^{11}$Be & $^{12}$Be & $({\rm n},\gamma)$ \\
$^8$B & $^8$Be & $\beta^+$  &
$^{11}$Be & $^{11}$B & $\beta^-$ \\
$^8$Be & $^4$He$^*$ & decay &
$^{11}$B & $^{12}$B & $({\rm n},\gamma)$ \\
\cline{1-3}
\multicolumn{3}{|c||}{helium burning} &
$^{11}$B & $^{12}$C & $({\rm p},\gamma)$, $({\rm d},{\rm n})$ \\
$^4$He & $^{12}$C & $(2\alpha,\gamma)$ &
$^{12}$Be & $^{12}$B & $\beta^-$ \\
$^8$Be & $^4$He & decay &
$^{12}$B & $^{12}$C & $\beta^-$, $({\rm p},{\rm n})$ \\
\hline
\end{tabular}
\end{center}
\end{table}

In Fig.~\ref{f1} the nuclear
reaction network for primordial nucleosynthesis in the Standard
Big Bang (SBB) is shown. Fig.~\ref{f2} shows the pp--chain dominating
in low--mass main--sequence stars, like our sun, whereas the CNO--cycle
takes over for more massive stars.
There are two other scenarios additionally
to the before discussed s-- and r--process where neutron capture plays
also a dominating
role. The first one is an alternative to the standard Big
Bang and is called the Inhomogenous Big Bang (IBB) with neutron
densities up to about $N_{\rm n} \simeq 10^{20}$\,cm$^{-3}$. The second
one is the so-called $\alpha$--process occuring
in type--II supernovae with neutron densities of about
$N_{\rm n} \simeq 10^{20-30}$\,cm$^{-3}$.
Fig.~\ref{f3} shows the net--work up to a mass number of
$A=12$ which should be taken into account additionally
to Fig.~\ref{f2} for primordial nucleosynthesis in the IBB
and the $\alpha$--process. These are neutron--rich
scenarios, where reactions involving light nuclei on the neutron--rich side of
the
region of stability have to be included.
The main nuclear reactions up to a mass number of $A=12$ taking place
in the  SBB, IBB, pp--chain, helium burning and the
$\alpha$--process are listed
in Table~\ref{t2}.

We want to discuss the $\alpha$--process a little more,
because the reaction rates calculated in the next section
occur in this scenario.
The $\alpha$--process takes place in the
neutrino--heated hot bubble between the nascent neutron star and
the overlying stellar mantle of a type--II supernova
\cite{woo92,mey92,how93,woo94,wit94}. During the
gravitational collapse in the supernova
the temperature in the core of the
star gets so high (up to $10^{10}$\,K)
that all the nuclei are disassembled through
photodissociation into protons and neutrons.
Afterwards, the material being
originally in nuclear statistical equilibrium (NSE) at high temperature,
is expanded and cooled so rapidly that not all the $\alpha$--particles
have time to reassemble. In this environment most
nucleons are either in the form of free neutrons or bound in
$\alpha$--particles.
The nucleosynthesis in the neutrino bubble takes place in subsequent steps:
first, $\alpha$--particles are formed from the free nucleons. Then, in the
following $\alpha$--process, nuclei up to about $A \approx 100$
are produced \cite{woo92}. Finally, the neutrino bubble is also an ideal site
for the r--process
synthesizing the elements of about $ A \ge 100$~\cite{woo92,kra93,tak94}.
\clearpage

\section{Few--Body Calculations}

Few--body methods have been used only partially for the calulation
of astropysical S--factors or reaction rates of
astrophysically relevant nuclear reactions. An example is an
exact three--nucleon calculation for radiative
capture by deuterons starting from realistic
NN--potentials.
In this calculation the astrophysically relevant
reaction rates for $^2$H(p,$\gamma$)$^{3}$He at thermonuclear energies and
$^2$H(n,$\gamma$)$^{3}$H at thermal energies
have been determined~\cite{fri91}.
Another example is the calculation of the
non--resonant part of the cross section for $^4$He(d,$\gamma$)$^{6}$Li
in the thermonuclear energy region. It was calculated
in an $\alpha$+p+n model by solving the three--body
Faddeev equations~\cite{blo93} and using a three--body
variational approach in the asymptotic region~\cite{kuk84}.

In the following we present the two--step processes $^4$He(2n,$\gamma$)$^6$He
and the photodisintegration of $^6$He as an example for an application
of few--body calculations to astrophysically relevant nuclear
reactions.
A more extensive presentation of this calculation
is given in ref.~\cite{efr95}.

The reaction rate per particle triplet for $^{4}$He(2n,$\gamma$)$^{6}$He, in
analogy to the triple--alpha process, is given by \cite{nom85}
\begin{eqnarray}\label{e1}
\left<2{\rm n}^4{\rm He}\right> & = & 2 \int_{0}^{\infty}dE_1\,
\frac{\hbar}{\Gamma\left(^5{\rm He},E_1\right)}
\frac{d\left<{\rm n}^4{\rm He}\right>\left(E_1\right)}{dE_1}\nonumber\\
&& \int_{0}^{\infty}dE_2\,
\frac{d\left<{\rm n}^5{\rm He}\right>\left(E_1,E_2\right)}{dE_2}
\quad ,
\end{eqnarray}
where $E_1$ and $E_2$ are the relative energies in the ($^{4}$He+n)--
and ($^{5}$He+n)--system, respectively. To determine
the reaction rate in Eq.~(\ref{e1}),
an integration over both energies $E_1$ and $E_2$ has to be
performed.
The quantity $\Gamma\left(^5{\rm He},E_1\right)$ is the energy--dependent width
of $^{5}$He,
whereas
$d\left<{\rm n}^4{\rm He}\right>\left(E_1\right)/dE_1$
and
$d\left<{\rm n}^5{\rm He}\right>\left(E_1,E_2\right)/dE_2$
are the integrands of the reaction rates $\left<\sigma v\right>$ \cite{rol88}
for the
first and second step, respectively.

To calculate the second step reaction rate $\left<{\rm n}^5{\rm He}\right>$,
the $^{5}$He(n,$\gamma$)$^{6}$He cross section is required.
One may estimate that the electric dipole transition E1 from the
incident s--wave is dominating the reaction. This is confirmed
by both three--body and direct--capture calculations. The cross section is
given by
\begin{equation}\label{e2}
\sigma_{\rm E1}\left(E_1,E_2\right) = \frac{2\pi}{81} \left(
\frac{E_\gamma}{\hbar c}
\right)^3 \frac{e^2}{\hbar v_2} \mid I \mid ^2
\end{equation}
with
\begin{equation}\label{e3}
I = \frac{1}{k_2} \int_0^\infty \!\! dr_1 \int_0^\infty \!\! dr_2
u_{\rm b}(E_1,r_1) r_1 u_{\rm sc} (E_2,r_2) r_2^2  f_{\ell=1,j=3/2} (r_1,r_2)
\quad ,
\end{equation}
where $E_\gamma=E_1+E_2+Q_{12}$ is the energy of the photon, with $Q_{12}$
being the
Q--value of the reaction $^{4}$He(2n,$\gamma$)$^{6}$He. In Eq.~(\ref{e3})
the coordinates
$r_1$ and $r_2$ are the distances of the valence neutrons from the
$\alpha$--core
and $^{5}$He, respectively.
The quantities $v_2$ and $k_2$ are the relative velocity and wave number in the
entrance channel
($^{5}$He+n) of the second step.
The quantity
$u_{\rm b}(E_1,r_1)/r_1$ represents the radial part of the
quasi--bound wave function of the $^{5}$He--resonance, and
$u_{\rm sc} (E_2,r_2)/r_2$ is the radial part of
the ($^{5}$He+n)--scattering wave function. The function
$f_{\ell=1,j=3/2} (r_1,r_2)$ is the radial component in the
expansion
\begin{equation}\label{e4}
\Psi\left(^6{\rm He}\right) = \sum_{\ell j} f_{\ell j}(r_1,r_2)
\left[\chi_{\ell j}\left({\hat{\bf r}}_1,\sigma_{1z}\right)
\chi_{\ell j}\left({\hat{\bf r}}_2,\sigma_{2z}\right)\right]_{J=0}
\end{equation}
of the $\alpha$+n+n ground--state wave function.
Here $\chi_{\ell j}$ are the angular--mo\-men\-tum functions
with orbital angular momentum $\ell$ and total
angular momentum $j=\ell \pm 1/2$, and the
brackets $[\ldots]$ denote vector coupling.
The expansion includes only equal $\ell$--values for
both neutrons and only the $(\ell=1,j=3/2)$--component
contributes to the cross section.

To calculate the first step reaction rate $\left<{\rm n}^4{\rm He}\right>$
we need the ($^{4}$He+n)--potential. It was obtained by fitting
a folding--type potential~\cite{kob84,ohu91}
to the experimental energy and width of the $^{5}$He ground state.
For the calculation of the energy dependent width
$\Gamma\left(^5{\rm He},E_1\right)$ of $^{5}$He in Eq.~(\ref{e1})
we used an equivalent square well potential with the same
r.m.s.~radius of the density distribution as for the folding potential.

In the second step the ground--state wave function $u_{\rm b}(E_1,r_1)$ of
$^{5}$He
was determined by solving the single--particle Schr\"odinger equation,
using the ($^{4}$He+n)--potential and a boundary condition for a decaying
system. It was then cut
at a reasonable cut--off radius.
For the determination of the scattering wave function $u_{\rm sc} (E_2,r_2)$ we
used again folding--type potentials.
An accurate three--body wave function of
$^{6}$He (see \cite{zhu93}) was used
in our calculations.

\begin{table}
\caption{\label{t3} Relevant input parameters for our calculations:
$Q_{12}$ is the Q--value for both steps, $J^{\Pi}$, $E_{\rm R}$ and
$\Gamma_{\rm R}$
are the spin--parity assignments, resonance energies and widths of the
indermediate nuclei $^5$He,
$^7$He and $^{10}$Li}
\begin{tabular}{crrrr}
\hline
reaction & $Q_{12}$\,[MeV] & $J^{\Pi}$ & $E_{\rm R}$\,[MeV]  &
$\Gamma_{\rm R}$\,[MeV] \\
\hline
$^4$He(2n,$\gamma$)$^6$He & 0.98$^{\rm a})$ & $\frac{3}{2}^{-}$$^{\rm b})$ &
0.89$^{\rm b}$) & $0.76 \pm 0.11$$^{\rm c}$) \\
$^6$He(2n,$\gamma$)$^8$He & 2.14$^{\rm a})$ &  $\frac{3}{2}^{-}$$^{\rm b})$ &
0.44$^{\rm b}$) & $0.16 \pm 0.03$$^{\rm b}$) \\
$^9$Li(2n,$\gamma$)$^{11}$Li & 0.31$^{\rm a})$ &  $1^{+}$$^{\rm d})$ &
0.54$^{\rm d}$) & $0.22 \pm 0.01 $$^{\rm d}$) \\
\hline
\end{tabular}\\
{\footnotesize $^{\rm a})$ Ref.~\cite{aud93};
$^{\rm b})$ Ref.~\cite{ajz88}; $^{\rm c})$ Ref.~\cite{efr95a}; $^{\rm d})$
Ref.~\cite{you94}}
\end{table}

The relevant input parameters for the calculations of the
reaction rate are shown
in Table~\ref{t3}. The calculated cross section for
$^{5}$He(n,$\gamma$)$^{6}$He
can be parametrized by
$\sigma = \sigma_{1}
/\left(E_2\,[{\rm {MeV}}]\right)^{1/2}$\,[$\mu$b].
The $^{5}$He(n,$\gamma$)$^{6}$He cross section exhibits the
well--known $1/v$--behavior for incident s--waves.

\begin{table}
\caption{\label{t4} Parameters of the cross section
for $^{5}$He(n,$\gamma$)$^{6}$He and the
reaction rates for $^{4}$He(2n,$\gamma$)$^{6}$He
and the reverse photodisintegration of $^{6}$He
calculated in our three--body model and in direct capture}
\begin{tabular}{c|rr|rr}
\hline
& \multicolumn{2}{c|}{Three--body calculation}
& \multicolumn{2}{c}{Direct--capture calculation}\\
\hline
$\sigma_{1}$\,[MeV$^{1/2}$\,$\mu$b] & \multicolumn{2}{c|}{0.152} &
\multicolumn{2}{c}{0.270}\\
\hline
& $0.1 \le T_9 \le 2$ & $2 < T_9 \le 15$ & $0.1 \le T_9 \le 2$ & $2 < T_9 \le
15$\\
$a$ & 0.00265 & 0.293 & 0.00471 & 0.520 \\
$b$ & 2.55 & $-$0.351 & 2.55 & $-$0.351 \\
$c$ & 0.181 & $-$5.24 & 0.181 & $-$5.24 \\
$d$ & 3.14 & 0.0286 & 3.14 & 0.0286 \\
$e$ & $-$1.05 & 1.85 & $-$1.05 & 1.85 \\
$f$ & $-$21.9 & $-$16.5 & $-$21.9 & $-$16.5 \\
\hline
\end{tabular}
\end{table}

The obtained reaction rate can be parametrized in
the following way:

\begin{equation}\label{e5}
N_{\rm A}^2 \left<2{\rm n}^4{\rm He}\right> = a
T_9^b \exp\left(\frac{c}{T_9}\right)\,
10^{-8}\,{\rm{cm}}^6\,{\rm s}^{-1}\,{\rm {mol}}^{-2}
\quad ,
\end{equation}
where the parameters $a,b,c$ are shown
in Table~\ref{t4} for the two temperature regions
$0.1 \le T_9 < 2$ and $2 \le T_9 \le 15$ ($T_9$: in units of
$10^9$\,K).
The inverse reaction rate $\lambda_\gamma$ per nucleus per second
can be calculated by using the RevRatio \cite{fow75}
\begin{equation}\label{e5a}
\lambda_\gamma={\rm RevRatio}\times N_{\rm A}^2 \left<2{\rm n}^4{\rm
He}\right>\,
{\rm s^{-1}}
\quad ,
\end{equation}
where we assumed that the photodisintegration proceeds through the $^5$He
resonance.
The RevRatio is parametrized in the following way
\begin{equation}\label{e5b}
{\rm RevRatio}=d T_9^e \exp\left(\frac{f}{T_9}\right)\, 10^{23}\,{\rm
cm^{-6}\,mol^{2}}
\quad .
\end{equation}
The parameters $d,e,f$ are also shown in Table~\ref{t4}.

The reaction rate
for $^{4}$He(2n,$\gamma$)$^{6}$He calculated with the
help of our three--body model (second column of Table~\ref{t4})
is at least more than three orders of magnitude larger than the
previously adopted value~\cite{fow75}. This is mainly due to the
non--resonant E1--transition of the second step $^{5}$He(n,$\gamma$)$^{6}$He,
which dominates the cross section, and which was not taken
into account previously.

\section{Direct--capture (DC) calculations}

We also calculated the reaction rates for $^4$He(2n,$\gamma$)$^6$He
and the reverse photodisintegration of $^6$He
by using the DC model for the second step
of the reaction. In this case Eq.~\ref{e3} reduces to

\begin{equation}\label{e6}
I = \frac{S_{\ell=1,j=3/2}}{k_2} \int_0^\infty \!\! dr
u_{\rm sc} (E_2,r) r^2  u_{\ell=1,j=3/2} (r) \quad ,
\end{equation}
where $r \equiv r_2$ and $S_{\ell=1,j=3/2}$ is the spectroscopic factor
and $u_{\ell=1,j=3/2}(r)$ is the wave function
for $^{6}$He = $^{5}$He+n.

The spectroscopic factor for $^5$He(n,$\gamma$)$^6$He
was calculated by shell--model calculations using
the (6--16) 2BME interaction of Cohen and Kurath~\cite{coh65}
and is given by $S_{\ell=1,j=3/2}= 1.3139$.
All the other parameters (see Table~\ref{t3}) for the DC calculation of
$^4$He(2n,$\gamma$)$^6$He were assumed to be the same as in the
three--body calculation. For the DC calculation we used
the code TEDCA~\cite{kra92}.

As can be seen from Table~\ref{t4} the
cross section for the second step and the total reaction rate of
$^4$He(2n,$\gamma$)$^6$He is enhanced by 78\% enhanced compared
to the three--body calculations.
Therefore, we assumed that the DC model is also
adequate within a factor of two to calculate the reaction
rates for two other two--step processes involving halo nuclei like
$^6$He(2n,$\gamma$)$^8$He and
$^9$Li(2n,$\gamma$), and the photodisintegration
of $^8$He and $^{11}$Li. These reactions
could also be of importance
in the $\alpha$--process.

We assume as for $^4$He(2n,$\gamma$)$^6$He
that again two p--neutrons are transferred in the
two--step process and that in the second step the E1--transition
is dominating. For $^9$Li(2n,$\gamma$)$^{11}$Li
we did not consider yet possible predicted low--energy intruder s--states in
$^{10}$Li~\cite{kry93,you94} and/or $^{11}$Li~\cite{tho94,shi95}.
Such states could lead to s--wave
resonances in $^{11}$Li near the ($^{10}$Li+n)--threshold
modifying our calculated non--resonant cross section
for $^{10}$Li(n,$\gamma$)$^{11}$Li. Furthermore,
by considering such s--wave states the two--step process could also proceed
over
$1^-$-- or $2^-$-- states in $^{10}$Li and not only
over the $1^+$--state at 0.54\,MeV above threshold (see Table~\ref{t3}) that
we considered in our DC--calculation.

The used input parameters are again
shown in Table~\ref{t3}. The spectroscopic factors
are given by $S_{\ell=1,j=3/2}=3.2047$ for $^7$He(n,$\gamma$)$^8$He,
and $S_{\ell=1,j=3/2} = 0.647$ and $S_{\ell=1,j=1/2} = 0.024$
for $^{10}$Li(n,$\gamma$)$^{11}$Li. For the
potentials in the exit channels we used the folding method
adjusted to the separation energy and the
r.m.s.~radii of the neutrons for the residual nuclei.

\begin{table}
\caption{\label{t5} Parameters for the cross section
of the second step and reaction rates of $^6$He(2n,$\gamma$)$^8$He and
$^9$Li(2n,$\gamma$)$^{11}$Li, and the reverse photodisintegration
of $^8$He and $^{11}$Li calculated in direct capture}
\begin{tabular}{c|rr|rr}
\hline
& \multicolumn{2}{c|}{$^6$He(2n,$\gamma$)$^8$He} &
\multicolumn{2}{c}{$^9$Li(2n,$\gamma$)$^{11}$Li}\\
\hline
$\sigma_{\rm 1\,MeV}$\,[$\mu$b] & \multicolumn{2}{c|}{1.828} &
\multicolumn{2}{c}{0.2435}\\
\hline
& $0.1 \le T_9 \le 2$ & $2 < T_9 \le 15$ & $0.1 \le T_9 \le 2$ & $2 < T_9 \le
15$\\
$a$ & 8.015 & $36.91$ & 0.651 & 0.956 \\
$b$ & $-$0.209 & $-$1.187 & 0.042 & $-$0.412 \\
$c$ & $-$2.514 & $-$4.262 & $-$3.394 & $-$3.432 \\
$d$ & 0.013 & 0.003 & 0.004 & 0.003 \\
$e$ & 1.709 & 2.687 & 1.458 & 1.912 \\
$f$ & $-$27.427 & $-$25.679 & $-$6.470 & $-$6.432 \\
\hline
\end{tabular}
\end{table}

The results for the parameters of the cross sections for
the second step and the reaction rates are also shown in Table~\ref{t5}. The
reaction rate for $^6$He(2n,$\gamma$)$^8$He is at $T_9 \approx 1$
about a factor 40 larger than for $^4$He(2n,$\gamma$)$^6$He.
Very similar results for the reaction rate
of $^4$He(2n,$\gamma$)$^6$He
have also been obtained in another DC calculation
by G\"orres et al.~\cite{gor95}.
However, our reaction rate for $^6$He(2n,$\gamma$)$^8$He
is more than a factor 10 larger than in ref.~\cite{gor95},
mainly be\-cause we used a larger spectroscopic factor.
The reaction rate for $^9$Li(2n,$\gamma$)$^{11}$Li (Table~\ref{t5})
is at $T_9 \approx 1$ almost a factor two enhanced compared
to $^4$He(2n,$\gamma$)$^6$He (Table~\ref{t4}),
even though the total Q--value for the
Li--reaction is a factor three smaller (Table~\ref{t3}).
This is mainly due to the more pronounced
halo structure of the Li--nuclei as compared to the
He--nuclei.

\begin{acknowledge}
The authors are indebted to I.J.~Thompson
for help in their work and to J.S. Vaagen and M.V. Zhukov for valuable
discussions on halo nuclei.
This work was supported by the
Fonds zur F\"orderung der wissenschaftlichen
For\-schung in \"Osterreich (project P10361--PHY), by the
\"Oster\-rei\-chi\-sche Nationalbank (project 5054) and by the International
Science Foundation (grant J4M100). Partial
financial support by the Russian--British--Northic Theory
collaboration (RNBT) is acknowledged.
\end{acknowledge}

\catcode`\@=11 \if@amssymbols%
\clearpage % forces LaTeX to finish all remaining open floats
\else\relax\fi\catcode`\@=12

\SaveFinalPage

\begin{thebibliography}{99}

\bibitem{rol88} C.E. Rolfs, W. Rodney: {\it Cauldrons in the Cosmos}.
Chicago: University of Chicago Press 1988

\bibitem{woo92} S.E. Woosley, R.D. Hofmann:
Astrophys. J. {\bf 395}, 202 (1992)

\bibitem{mey92} B.S. Meyer et al.:
Astrophys. J. {\bf 399}, 656 (1992)

\bibitem{how93} W.M. Howard et al.:
Astrophys. J. {\bf 417}, 713 (1993)

\bibitem{woo94} S.E. Woosely et al.:
Astrophys. J. {\bf 433}, 229 (1994)

\bibitem{wit94} J. Witti, H.-Th. Janka and K. Takahashi:
Astron. and Astrophys. {\bf 286}, 841 (1994)

\bibitem{kra93} K.L. Kratz et al.:
Astrophys. J. {\bf 403}, 216 (1993)

\bibitem{tak94} K. Takahashi, J. Witti and H.-Th. Janka:
Astron. and Astrophys. {\bf 286}, 857 (1994)

\bibitem{fri91} J.L. Friar et al.: Phys. Rev. Lett. {\bf 66}, 1827 (1991)

\bibitem{blo93} L.D. Blokhintsev et al.:
Phys. Rev. C {\bf 48}, 2390 (1993)

\bibitem{kuk84} V.I. Kukulin et al.:
Nucl. Phys. A {\bf 417}, 128 (1984); {\bf 453}, 365 (1986)

\bibitem{efr95} V.D. Efros et al.:
Phys. Lett. B (submitted)

\bibitem{nom85} K. Nomoto, F.-K. Thielemann, S. Hiyaji:
Astron. and Astrophys. {\bf 149}, 239 (1985)

\bibitem{kob84} A.M. Kobos et al.:
Nucl. Phys. A {\bf 425}, 205 (1984)

\bibitem{ohu91} H. Oberhummer, G. Staudt: In: H. Oberhummer,
(ed.): {\it Nuclei in the Cosmos\/}, p.29. Heidelberg: Springer Verlag 1991

\bibitem{zhu93} M.V. Zhukov et al.:
Physics Rep. {\bf 231}, 151 (1993)

\bibitem{aud93} G. Audi, A.H. Wapstra: Nucl. Phys. A {\bf 565}, 1 (1993)

\bibitem{ajz88} F. Ajzenberg--Selove: Nucl. Phys. A {\bf 490}, 1 (1988)

\bibitem{efr95a} V.D. Efros et al.:
Phys. Rev. C (to be submitted)

\bibitem{you94} B.M. Young et al.:
Phys. Rev. C {\bf 49}, 279 (1994)

\bibitem{fow75} W.A. Fowler, G.R. Caughlan and B. Zimmerman:
Ann. Rev. Astron. Astrophys. {\bf 13}, 69 (1975)

\bibitem{coh65} S. Cohen, D. Kurath: Nucl. Phys. {\bf 73}, 1 (1965)

\bibitem{kra92} H. Krauss: Computer code TEDCA. Technical University
Vienna, 1992

\bibitem{kry93} R. Kryger et al.:
Phys. Rev. C {\bf 47}, R4239 (1993)

\bibitem{tho94} I.J. Thompson, M.V. Zhukov:
Phys. Rev. C {\bf 49}, 1904 (1994)

\bibitem{shi95} S. Shimumora et al.:
Phys. Lett. B {\bf 348}, 29 (1995)

\bibitem{gor95} J. G\"orres et al.:
Phys. Rev. C (submitted and private communication)

\end{thebibliography}
\end{document}